\documentclass{elsart}

\usepackage{epsfig}
\usepackage{bm}
\usepackage{a4wide}

\usepackage{xspace}

\newcommand{\GeV}{\ensuremath{\mbox{GeV}}\xspace}
\newcommand{\GeVc}{\ensuremath{\mbox{GeV}/c}\xspace}
\newcommand{\GeVcc}{\ensuremath{\mbox{GeV}/c^2}\xspace}
\newcommand{\GeVVcc}{\ensuremath{\mbox{GeV}^2/c^2}\xspace}

\newcommand{\cm}{\ensuremath{\mbox{cm}}\xspace}

\newcommand{\epsp}{\ensuremath{\epsilon_\mathrm{P}}\xspace}
\newcommand{\pt}{\ensuremath{p_\mathrm{T}}\xspace}
\newcommand{\Esh}{\ensuremath{E_\mathrm{sh}}\xspace}
\newcommand{\mc}{\ensuremath{m_\mathrm{c}}\xspace}
\newcommand{\Bmu}{\ensuremath{B_\mathrm{\mu}}\xspace}

\begin{document}
\begin{frontmatter}

\title{Leading Order Analysis of Neutrino Induced Dimuon Events in the CHORUS Experiment}

\author{The CHORUS Collaboration}
\address{\mbox{~}}
\maketitle

\begin{abstract}
We present a leading order QCD analysis of a sample of neutrino induced charged-current 
events with two muons in the final state originating in the lead-scintillating fibre 
calorimeter of the CHORUS detector.
The results are based on a sample of 8910 neutrino and 430 antineutrino
 induced opposite-sign dimuon events collected during the exposure of
 the detector to the CERN Wide Band Neutrino Beam 
between 1995 and 1998.
The analysis yields a value of the charm quark mass of $\mc = (1.26\pm 0.16 \pm 0.09)\ \GeVcc $ and 
a value of the ratio of the strange to non-strange sea in the nucleon of $\kappa = 0.33 \pm 0.05 \pm 0.05$,
improving the results obtained in similar analyses by previous
 experiments.
\end{abstract}




\author{A.~Kayis-Topaksu, G.~\"{O}neng\"ut}

\address{{\bf \c{C}ukurova University, Adana, Turkey}}

\author{R.~van Dantzig,  M.~de Jong, R.G.C.~Oldeman$^1$}

\address{\bf NIKHEF, Amsterdam, The Netherlands}

\author{M.~G\"uler, S.~Kama,   U.~K\"ose, M.~Serin-Zeyrek,    P.~Tolun}

\address{\bf  METU, Ankara, Turkey}

\author{M.G.~Catanesi, 
M.T.~Muciaccia}

\address{\bf Universit\`a di Bari and INFN, Bari, Italy}

\author{A.~B\"ulte,         
K.~Winter}

\address{\bf  Humboldt Universit\"at, Berlin, Germany$^{2}$}

\author{B.~Van de Vyver$^{3,4}$, P.~Vilain$^{5}$, G.~Wilquet$^{5}$}

\address{\bf Inter-University Institute for High Energies (ULB-VUB) Brussels, Belgium}

\author{B.~Saitta}

\address{\bf Universit\`a di Cagliari and INFN, Cagliari, Italy}

\author{E.~Di Capua}

\address{\bf Universit\`a di Ferrara and INFN, Ferrara, Italy}

\author{S.~Ogawa, H.~Shibuya}

\address{\bf Toho University,  Funabashi, Japan}

\author{I.R.~Hristova$^6$,  T.~Kawamura, D.~ Kolev$^7$, M.~Litmaath, H.~ Meinhard,}
\author{J.~Panman, A.~Rozanov$^{8}$, R.~Tsenov$^{7}$, J.W.E. Uiterwijk, P. Zucchelli$^{3,9}$}

\address{\bf CERN, Geneva, Switzerland}

\author{J.~Goldberg}

\address{\bf Technion, Haifa, Israel}

\author{M.~Chikawa}

\address{\bf Kinki University, Higashiosaka, Japan}

\author{J.S.~Song, C.S.~Yoon}

\address{\bf Gyeongsang National University,  Jinju, Korea}

\author{K.~Kodama, N.~Ushida}

\address{\bf Aichi University of Education, Kariya, Japan}

\author{S.~Aoki, T.~Hara}

\address{\bf Kobe University,  Kobe, Japan}

\author{T.~Delbar,  D.~Favart, G.~Gr\'egoire, S.~ Kalinin, I.~ Makhlioueva}

\address{\bf Universit\'e Catholique de Louvain, Louvain-la-Neuve, Belgium} 

\author{A.~Artamonov, P.~Gorbunov, V.~Khovansky, V.~Shamanov, I.~Tsukerman}

\address{\bf Institute for Theoretical and Experimental Physics, Moscow, Russian
Federation}

\author{N.~Bruski, D.~Frekers,
D.~Rondeshagen, T.~Wolff}      

\address{\bf Westf\"alische Wilhelms-Universit\"at, M\"unster, Germany$^{2}$}

\author{K.~Hoshino, 
J.~Kawada, 
M.~Komatsu,
M.~Miyanishi, 
M.~Nakamura,}
\author{ T.~Nakano, K.~Narita, K.~Niu, K.~Niwa, 
N.~Nonaka, O.~Sato, T.~Toshito}

\address{\bf Nagoya University, Nagoya, Japan}

\author{S.~Buontempo, A.G.~Cocco, N.~D'Ambrosio,
G.~De Lellis, G.~ De Rosa,}
\author{F.~Di Capua, 
G.~Fiorillo, A.~Marotta,
M.~Messina, P.~ Migliozzi,} 
\author{R.~Santorelli,                
L.~Scotto Lavina, 
P.~ Strolin, V.~Tioukov}

\address{\bf Universit\`a Federico II and INFN, Naples, Italy}

\author{T.~Okusawa}

\address{\bf Osaka City University, Osaka, Japan}

\author{U.~Dore, P.F.~Loverre,
L.~Ludovici, 
G.~Rosa, R.~Santacesaria,}
\author{A.~Satta, F.R.~Spada}

\address{\bf Universit\`a La Sapienza and INFN, Rome, Italy}

\author{E.~Barbuto, C.~Bozza, G.~Grella, G.~Romano, C.~Sirignano,}
\author{S.~Sorrentino}

\address{\bf  Universit\`a di Salerno and INFN, Salerno, Italy}

\author{Y.~Sato, I.~Tezuka}

\address{\bf Utsunomiya University,  Utsunomiya, Japan}

{\footnotesize
---------

\begin{flushleft}

$^{1}$ Now at University of Cagliari, Cagliari, Italy.
\newline
$^{2}$ {Supported by the German Bundesministerium f\"ur Bildung und Forschung under contract numbers 05 6BU11P and 05
7MS12P.}
\newline
$^{3}$ {Now at SpinX Technologies, Geneva, Switzerland.}
\newline
$^{4}$ {Fonds voor Wetenschappelijk Onderzoek, Belgium.}
\newline
$^{5}$ {Fonds National de la Recherche Scientifique, Belgium.}
\newline
$^{6}$ {Now at DESY, Hamburg.}
\newline
$^{7}$ {On leave of absence and at St. Kliment Ohridski University of Sofia, Bulgaria.}
\newline
$^{8}$ {Now at CPPM CNRS-IN2P3, Marseille, France.}
\newline
$^{9}$ {On leave of absence from INFN, Ferrara, Italy.}
\end{flushleft}
}


\end{frontmatter}


\section*{Introduction}

In neutrino-nucleon deep-inelastic scattering, events that present two muons in the final
state are mainly due to the muonic decay of a charmed hadron produced in
a neutrino charged-current interaction. 
The charm quark can be produced both through 
$d\rightarrow c$ and $s\rightarrow c$ weak currents. The $\bar{s}\rightarrow \bar{c}$
transition dominates in the antineutrino induced dimuon events, whilst in neutrino
induced interactions the $d\rightarrow c$ and $s\rightarrow c$ transition have comparable
contributions, as the large valence $d$ quark content of the nucleon
compensates for the Cabibbo
suppressed $d\rightarrow c$ transition.
Dimuon events in neutrino interactions can be used to measure the
strange quark content of the nucleon and the value of the charm quark mass.
Moreover, the charm production mechanism is of great importance for testing 
perturbative QCD predictions.

Neutrino charm production has been investigated with counter experiments
like CDHS~\cite{cdhs}, CCFR~\cite{ccfr}, CHARM II~\cite{charmii}, NOMAD~\cite{nomad}
and NuTeV~\cite{nutev}, as well as bubble chambers (BEBC~\cite{bebc})
and nuclear emulsion detectors like E531~\cite{e531} and
CHORUS~\cite{plb604,plb613,sergei}.
In this paper we present the study of a sample of neutrino induced charged-current
dimuon events produced in the lead-scintillating fibre calorimeter of the CHORUS detector;
the collected statistics is the second largest to date. 
The analysis is performed in the framework of the leading order QCD
formalism and the result
will be expressed in terms of the strange content of the nucleus $\kappa$, 
the charm quark mass $\mc$, the Peterson fragmentation parameter \epsp
and the charmed hadron semileptonic branching ratio $\Bmu$.

\section{The CHORUS experiment} 

The CHORUS experiment has been designed for a short-baseline search of neutrino oscillations
in the $\nu_\mu$-$\nu_\tau$ channel at relatively high $\Delta m^2$~\cite{chorus}. 
The detector was exposed to the Wide Band Neutrino Beam 
at CERN from 1995 until 1998. Neutrinos were produced at the SPS by a proton beam of
450~GeV; the average $\nu_\mu$ energy was 27~GeV with a $\bar{\nu}_\mu$ contamination
of about 6\%. The detector is described in detail in Ref.~\cite{chorusdet}; it consists of
a nuclear emulsion target, a fibre tracker, an air gap hadron spectrometer, a lead-scintillating 
fibre calorimeter and a muon spectrometer.

The 112 ton lead-scintillating fibre calorimeter, described in detail
in Ref.~\cite{calor}, consists of three sections (EM, HAD1, HAD2) with decreasing granularity along
the beam direction, for a total of 5.2 interaction lengths. 
Owing to its high mass,
the calorimeter also provides an active target for incoming neutrinos.
A large number of neutrino charged-current interactions on lead was 
collected during the four years of data taking.

A test beam calibration of the calorimeter performances was carried out
during the data taking and results are reported in Ref.~\cite{calor}.
Two neural net algorithms were developed in order to improve the vertex 
position resolution and the energy resolution for charged-current events originating in the 
calorimeter.
Both algorithms have been tuned for the purposes of this analysis on
Monte-Carlo neutrino induced events which present more than one reconstructed muon in the final state.
The vertex finding algorithm achieved a longitudinal and transverse position resolution of
$\sigma(V_x)=4.0$ cm and $\sigma(V_{y,z})= 3.0$~cm, respectively, with an average efficiency of
finding the correct calorimeter plane of (82 $\pm$ 1)\%. The energy resolution was estimated to
be $\sigma(E_\nu)/E_\nu = 0.39/\sqrt{E_\nu} + 0.14$ (with $E_\nu$
expressed in GeV).

A muon spectrometer composed of six magnetized iron toroids instrumented with drift chambers
and scintillators is located further downstream to identify muons and determine their 
momentum.
A resolution varying from $\Delta p/p \sim 0.15$ at 20~\GeVc to $\Delta p/p 
\sim 0.19$ at 70~\GeVc was achieved, as measured with test beam muons.

The trigger system of the CHORUS experiment is described in
Ref.~\cite{trigger}.
A dedicated muon trigger was set-up to collect a large statistic of ``double track'' events
used both in dimuon and trimuon analyses~\cite{trimu}.
A total of $4.9\times 10^{19}$ protons were accumulated on the Be target in the period 
1995--1998 and the overall CHORUS data collection efficiency was about 92\% with a
dead time of 5\%.

\section{Theoretical framework}

\label{framew}

Neutrino interactions with two muons in the final state mainly occur
when the weak charged-current on a $d$ or $s$ quark produces a charmed quark and
the charmed hadron subsequently decays 
with a probability $\Bmu$ into a muon and other hadrons. 
The primary muon and the decay muon have opposite electric charge.
The large value of the charm quark mass gives rise to an energy threshold 
for the process; this is described in the leading order QCD formalism
by the so called ``slow rescaling'' mechanism~\cite{slowr}.

\subsection{Leading order dimuon cross-section}

The leading order (LO) QCD framework of deep-inelastic neutrino scattering (DIS)
has been adopted to describe charm quark production. This scheme is 
reported in Ref.~\cite{aivazis} and uses the 
helicity formalism to describe the neutrino charged-current
cross-section.
It has the advantage of treating in a natural way the
different mass scales involved in the heavy quark production process.
The neutrino charm production cross-section is written as
\begin{equation}
\frac{d^2 \sigma}{dx dy} = 2G_F^2 \frac{yQ^2}{\pi} \left( |V_\mathrm{cd}|^2 d(\chi) + 
|V_\mathrm{cs}|^2 s(\chi) \right) \left[ \left( \frac{1+\cosh \psi}{2}\right)^2 + 
\frac{\mc^2}{2Q^2}\frac{\sinh^2 \psi}{2} \right] \
\label{cslo}
\end{equation}
in terms of the Bjorken variables $x$ and $y$ and the Fermi constant $G_F$.
$V_\mathrm{cd}$ and $V_\mathrm{cs}$ are the CKM matrix elements, $d(x)$ and $s(x)$ being the
momentum distributions of the scattered quarks.
The angle $\psi$ is related to $E_\nu$, $E_\mu$ and $Q$ ,the incoming neutrino energy, the 
leading muon energy and the $W$ boson four-momentum, respectively:
\begin{equation}
\cosh \psi = \frac{E_\nu+E_\mu}{\sqrt{Q^2+(E_\nu-E_\mu)^2}} \ .
\end{equation}
This reduces to $\frac{2-y}{y}$ when $\mc$ vanishes.
With $m_q$ the mass of the scattered quark and $M$ the nucleon mass, the scaling variable $\chi$ is defined as
\begin{equation}
\chi = \eta \frac{(Q^2 - m_q^2 + \mc^2) +
 \Delta(-Q^2,m_q^2,\mc^2)}{2Q^2} \ ,
\end{equation}
where
\begin{equation}
\frac{1}{\eta} = \frac{1}{2x} + \sqrt{\frac{1}{4x^2} + \frac{M^2}{Q^2}}
\end{equation}
and the function  $\Delta$ is given by
\begin{equation}
\Delta(a,b,c) = \left(a^2+b^2+c^2-2(ab+bc+ca) \right)^{\frac{1}{2}}.
\end{equation}
In the limit $M^2/Q^2\rightarrow 0$ the variable $\chi$ becomes the
usual ``slow rescaling'' variable~\cite{barnett}. 

When dealing with experimentally measured quantities, 
the expression~(\ref{cslo}) has to be corrected for the electromagnetic 
radiative processes arising in the final state whose effect
is to lower the primary muon energy and, correspondingly, to raise the hadronic 
shower energy. A correction for such an effect has been applied according to the
prescription of Bardin~\cite{bardin}.

\subsection{Parton distributions}

To interpret the experimental data in terms of charm quark production 
described by Eq.~(\ref{cslo}) a parametrization of the quark content
in the target nuclei is required. 
The GRV94LO parton distribution functions derived from
experimental deep-inelastic scattering data has been used~\cite{grv94lo}. 
To take into account the
non-isoscalarity of the lead-scintillating fibre calorimeter, the
total valence quark content has been parametrized as
\begin{equation}
x\ q_{\rm val}(x,Q^2) = \frac{A-Z}{A}\ x\ u_{\rm val}\ + \ \frac{Z}{A}\ x\ d_{\rm val} \ ,
\end{equation}
where, due to strong isospin symmetry, the
{\it down} quark content of the neutron is taken to be the same as the {\it up} quark content 
of the proton. The strange quark content in protons and neutrons is assumed to be the same.
The non-strange sea-quark contents of the nucleon is assumed 
to be symmetric between up and down quarks.
The strange quark content of the nucleon is described by the parameter
\begin{equation}
\kappa = \frac{\int \left[ xs(x,\mu_o^2) + x\bar{s}(x,\mu_o^2) \right] dx}{\int 
\left[ x\bar{u}(x,\mu_o^2) + x\bar{d}(x,\mu_o^2) \right] dx} \ ,
\end{equation}
where $\mu_o^2$ is an arbitrary reference scale chosen to be 20 GeV$^2$. 
In the following, an SU(3) flavour symmetric sea is assumed; this means that 
the strange and the anti-strange content of
the nucleon are the same and have the same $x$ dependence.

An additional parameter $\alpha$ is often introduced in similar analyses that
allows the strange quark to behave differently from the up and 
down quarks.
This is obtained by multiplying the strange quark 
parton distribution function ({\it pdf}) by
the factor $(1-x)^\alpha$. In the following the weighted average $\alpha=2 \pm 1$ of 
the values given by the CHARM II and CCFR leading order analyses is used.

\subsection{Charm quark fragmentation and meson decays}

The non-perturbative processes that act to ``dress'' the bare charm quark
produced in the neutrino interaction are known as fragmentation or hadronization.
The fragmentation is usually described as a function of the variable $z$
defined as the fraction of the longitudinal momentum taken over by the charm
hadron $h_\mathrm{c}$:
\begin{equation}
z=P_{L}(h_c)/P_{L}^\mathrm{max}(h_c) \ , 
\label{zdefined}
\end{equation}
where $P_\mathrm{L}^\mathrm{max}(h_\mathrm{c})$ is the maximum hadron longitudinal 
momentum  relative to the W-boson direction in the boson-nucleon center of 
mass reference frame. 
The Peterson parametrization of the charm quark fragmentation is written as ~\cite{peterson} 
\begin{equation}
D(z,\epsp) \propto z^{-1} \left( 1 - \frac{1}{z} -
				\frac{\epsp}{1-z} \right)^{-2} \ , 
\label{peters}
\end{equation}
where \epsp is a free parameter to be determined from the data.
In leading order QCD, the neutrino cross-section can be factorized as
\begin{equation}
\frac{{\rm d}^3 \sigma_\nu}{{\rm d}x{\rm d}y{\rm d}z} (\nu N \rightarrow \mu h_c X) =
\frac{{\rm d}^3 \sigma_\nu^{LO}}{{\rm d}x{\rm d}y{\rm d}z}
(\nu N \rightarrow \mu c X) D(z) \ .
\label{loexp}
\end{equation}
This is no longer true in the next-to-leading order formalism~\cite{nlo}.
The transverse momentum of charmed hadrons with respect to the W-boson direction 
is on average small ($\langle \pt^2 \rangle \sim 0.2$ \GeVVcc) and is assumed 
to be distributed as $dN/d \pt^2 \propto e^{-b \pt^2}$. The parameter $b$ is taken to be 
$1.1\ \GeVVcc$ following Ref.~\cite{lebc}.

The fragmentation variable $z$ cannot be determined on an
event-by-event basis since the direction of the charm quark or charmed hadron 
cannot be measured directly. The visible fragmentation variable is 
thus defined as 
\begin{equation}
z_{\rm vis} = \frac{E_{\mu 2}}{(E_{\mu 2}+E_{\rm sh})} \ ,
\label{zvis}
\end{equation}
where $E_{\mu 2}$ and $E_{\rm sh}$ are the energy of the muon
coming from the charmed hadron decay and the hadronic shower final state energy,
respectively. 
The relation between $z$ and $z_{\rm vis}$ has been studied with the aid 
of Monte-Carlo simulations.

\section{Data sample and event selection}

The dimuon sample was collected during 
four years of data taking using a dedicated trigger
setup based on the energy released in the fibre calorimeter and on the 
presence of two or more well separated tracks in the spectrometer~\cite{trigger}.
A total of $6.6 \times 10^6$ events tagged as dimuons by the
trigger were reduced to about $5.0 \times 10^5$ requiring that at least two muons
were well reconstructed in the muon spectrometer.
For each event, the reconstructed muon energy $E_\mu$, the angle of the muon with respect
to the neutrino beam direction $\theta_\mu$ evaluated at the interaction vertex
and the hadron shower energy $\Esh$ are used to derive the following kinematic variables:
\begin{itemize}
\item $E_\nu^{\rm vis} = \Esh + E_{\mu 1} +  E_{\mu 2}$, the visible neutrino energy
\item $Q^2_{\rm vis} = 2E_\nu^{\rm vis} E_{\mu 1}(1 - \cos \theta_{\mu 1}) - m_\mu^2$, the visible
      four-momentum transfer squared
\item $x_{\rm vis} = Q^2_{\rm vis}/2M(\Esh+ E_{\mu 2})$, the visible Bjorken $x$
\item $z_{\rm vis} = E_{\mu 2}/(\Esh + E_{\mu 2})$, the visible fragmentation variable, as reported
in Eq.~\ref{zvis}.
\end{itemize}
The final sample of dimuon events was selected imposing the following selection
criteria:
\begin{itemize}
\item[C1:] at least two muons must be well reconstructed, i.e. a minimum of five measured track
      points in the drift chambers placed in the gaps between the iron magnets
      is required for both muons;
\item[C2:] the longitudinal position of the event vertex ($V_x$) determined by the 
      Neural Net algorithm must be inside a fiducial
      volume given by $290.7~\cm < V_x < 384.0$~cm (this corresponds mainly to the
      HAD1 part of the calorimeter); the transverse vertex positions ($V_y$ and $V_z$)
      are required to satisfy $-120~\cm < V_{y,z} < 120$~cm to ensure hadronic
      shower lateral containment;
\item[C3:] the shower energy must be in the range
      $5~\GeV < \Esh\ < 150$~\GeV;
\item[C4:] to ensure a good reconstruction quality and control of the
      acceptances, a cut $E_\mu > 5$ GeV when extrapolated to 
      the vertex position in the calorimeter 
      is applied.  This cut also reduces the meson decay background;
\item[C5:] the reconstructed neutrino energy must be
      in the range $10~\GeV < E_{\nu} < 240$~GeV to ensure a good control of the detection 
	efficiencies;
\item[C6:] a cut $Q^2 > 3.0$ \GeVVcc is applied in order 
      to exclude regions in which the Monte-Carlo simulation is not reliable, 
      as described in the following section;
\item[C7:] the transverse distance between the two muons extrapolated to the interaction
      vertex $x$ coordinate ($d_{12}$) is requested to be less than $15$~cm to reject
      background due to hadron decays.
\end{itemize}
The primary muon is assumed to be the one with the highest \pt with respect
to the neutrino beam direction; this leads to
a (95.8 $\pm$ 1.0)\% efficiency to correctly identify the primary muon for
neutrino induced interactions and (94.0 $\pm$ 1.1)\% for antineutrino interactions.
In cases in which there are more than two reconstructed muons, the two most 
energetic ones 
are used in the analysis. 
The average charm induced opposite sign dimuon trigger efficiency
was evaluated by means of Monte-Carlo simulation on a sample of
events surviving the selection cuts and found to be (91 $\pm$ 6)\%.
The event statistics is reported in Table~\ref{esta}
where the top line indicates the charges of the primary and secondary muons respectively.

\begin{table}
\caption{\label{esta} Statistics for opposite-sign and same-sign dimuon events 
induced by neutrino interactions. The cuts are explained in the text.}
\begin{center}
\begin{tabular}{l ccccc}
Cut & $-+$ & $--$ & $+-$ & $++$ & total \\
\hline
C1 & 33251 & 30750 & 5769 & 899 & 70669 \\
C2 & 17517 & 12387 & 2144 & 203 & 32251 \\
C3 & 16894 & 11510 & 2023 & 186 & 30613 \\
C4 & 13948 & 10078 & 1689 & 141 & 25856 \\
C5 & 13938 & 10051 & 1688 & 140 & 25817 \\
C6 & 12579 &  8962 & 1270 & 110 & 22921 \\
C7 & 10218 &  1441 &  975 &  46 & 12680 \\
\end{tabular}
\end{center}
\end{table}

\begin{figure}
\centering \epsfig{figure=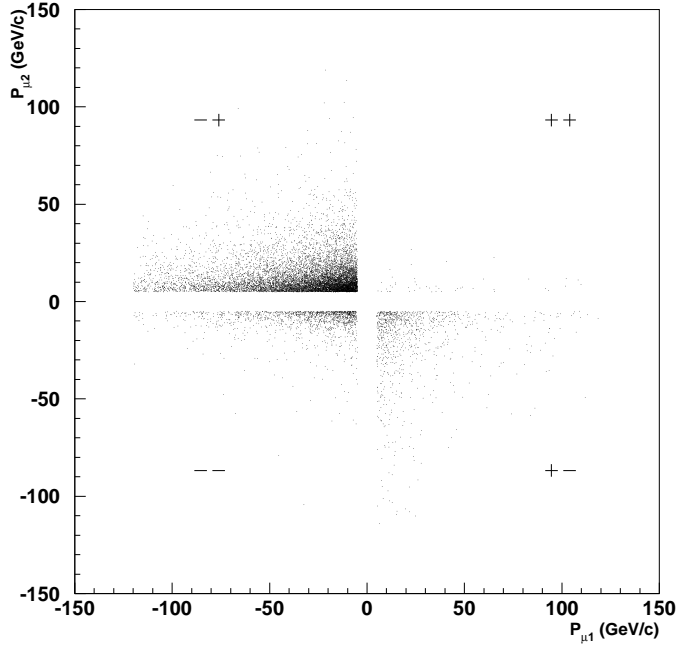,width=0.6\linewidth}
\caption{\label{data1} Scatter plot in the ($p_{\mu_1}$ ,$p_{\mu_2}$) plane of the dimuon selected events.
The sign of the muon momentum represents its electric charge.}
\end{figure}

The dimuon sample statistics after the selection is therefore
\begin{equation}
N^{-+} = 10218 \ ; \ N^{--} = 1441 \ ; \  N^{+-} = 975 \ ; \ N^{++} = 46 \ .
\end{equation}
Figure~\ref{data1} shows a scatter plot of the selected events in the
($p_{\mu1}$, $p_{\mu2}$)
plane.

\section{Monte-Carlo simulation}

A full simulation of the events induced by neutrinos with the production of a charmed quark
has been performed. 
The first step of the simulation consists in
the implementation of the formalism described in section~\ref{framew}.
A specific event generator has been developed by the CHORUS 
collaboration.
It implements JETSET~\cite{jetset} for 
charm quark fragmentation and charmed hadron decays following Eq.~(\ref{peters}).
The incoming neutrino beam is described according to a
GEANT 3.21~\cite{geant} based
package that describes the interaction of the SPS proton beam with the
beryllium target. 
All relevant components of the beam were part of the simulation.
The parton distribution functions were implemented with the
use of the PDFLIB~\cite{pdflib} package. 
The GRV94LO~\cite{grv94lo} {\it pdf}\,'s were adopted in the production of the 
sample of events used in the fit procedure.
The CTEQ3L~\cite{cteq3l} {\it pdf}\,'s have been used to obtain an estimate of the systematic error on the
final result. The value of the CKM mixing matrix elements has been taken from the PDG 
compilation~\cite{pdg}.
The maximum allowed incoming neutrino energy was $E_\nu^{\rm max}=300$ GeV and the kinematical bounds 
in the event generator were $Q^2_{\rm min}=2.0$~\GeVVcc and $W^2_{\rm min}=4.0\ {\rm GeV}^2$
in order to avoid quasi-elastic and resonance production processes.

The generator output is then used as input for
the CHORUS event simulation program.
Based on GEANT~3.21~\cite{geant}, it
contains the complete description of the apparatus and provides the
response of the various instrumented parts of the detector.
The events produced in this way undergo all the analysis steps performed 
for the real data.

\subsection{Background evaluation}

\begin{figure}
\centering \epsfig{figure=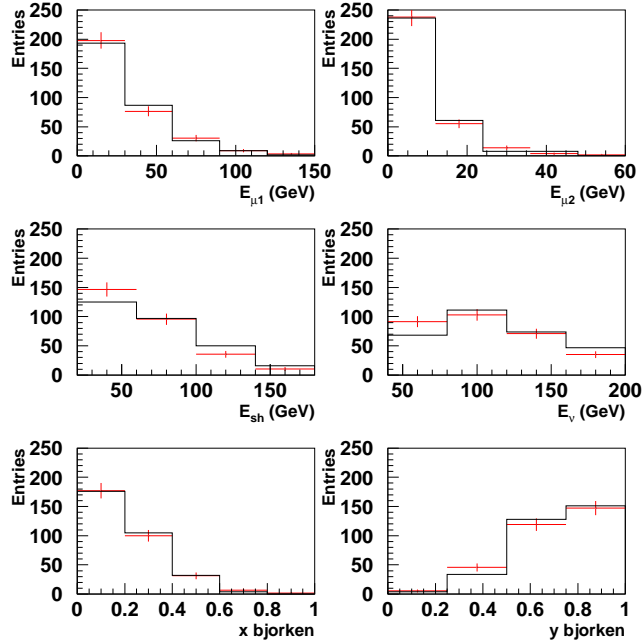,width=0.6\linewidth}
\caption{\label{ssmc} Comparison between reconstructed quantities for same-sign dimuon events 
in data (points) and MC sample (histogram).}
\end{figure}

The main source of background to the charm induced dimuon events is 
given by the muonic decay of non-charmed hadrons
produced either directly or during the shower development of a charged-current
event. The background level depends on the probability for a $\pi^\pm$ or 
a $K^\pm$ to decay before it interacts in the calorimeter or in the first
part of the muon spectrometer. The relative rates of same-sign and opposite-sign events
resulting from this background are closely related to the probability of producing positive
or negative pions and kaons during the hadronic shower development. 
A standard way to estimate 
its contribution to the opposite-sign dimuon sample relies on the admitted hypothesis that the 
same-sign dimuon events belong to the background. 
For instance, the background to the charm
induced opposite-sign dimuon events with a leading $\mu^-$ is given by 
\begin{equation}
B^{-+} = \left( \frac{N_{\rm CC}^{-+}}{N_{\rm CC}^{--}} \right)_{\rm MC} \times N_{\rm DATA}^{--} \ .
\label{bsub}
\end{equation}
where the charm to muon decay process has been disabled in the simulation of the 
neutrino induced charged-current events. 
A total of about 300 events were obtained after selection cuts out of $3\times10^6$
neutrino CC interactions for which a complete MonteCarlo simulation of the detector
response has been performed.
The comparison between same-sign neutrino dimuon events in data and MonteCarlo is shown in
Figure~\ref{ssmc}; the fairly good agreement between data and MonteCarlo validates the use of
relation~(\ref{bsub}).
Using this sample of events, a value of 
$(N_{\rm CC}^{-+}/N_{\rm CC}^{--})_{\rm MC} = 0.89 \pm 0.10$ has been
established. Once selection cuts have been applied, the number of events with
opposite sign dimuons is thus smaller than that of same sign
muons\footnote{The ratios between positive and negative pions and
kaons produced in the MonteCarlo sample of neutrino induced charged current
interactions are $\pi^+/\pi^-\sim 1.7$ and $K^+/K^-\sim 1.6$ for events having
$E_{\mu 1}>10$ GeV and $E_{\pi,K}>5$ GeV}; this is mainly
due to the fact that the magnet polarity of the muon spectrometer was chosen
to focus negative charged muons. Consequently, the efficiency to reconstruct a $\mu^+$
is smaller than to reconstruct a $\mu^-$.  The majority of background events is in the
region of small muon energy where the effect is more pronounced.
Similarly the value $(N_{\rm CC}^{+-}/N_{\rm CC}^{--})_{\rm MC} = 0.11
\pm 0.02$ is obtained and used to evaluate the non-charm  background to
the opposite-sign dimuon sample with a leading $\mu^+$.

\section{Results}

The number of observed dimuon events depends on the charm quark mass ($\mc$) via the slow rescaling
mechanism, the amount of strange quark sea ($\kappa$), the fragmentation parameter (\epsp)
and on the branching ratio of charm into muon ($\Bmu$).
From the initial sample of Table~\ref{esta}, taking into account selection efficiencies and 
neutrino--antineutrino 
cross contamination, a total of $8910 \pm 180$ events with a leading $\mu^-$ and 
$430 \pm 60$ events with a leading $\mu^+$ were selected.
This represents the second largest sample of neutrino induced dimuons to date.

\subsection{Maximum Likelihood four-parameter fit}

To extract information on the production mechanism of dimuon events, a maximum
likelihood fit of the Monte-Carlo events to the data
has been performed.
\begin{table}
\caption{\label{fitres} Results of the four parameter fit procedure and parameter correlation matrix 
given by MINUIT.}
\begin{center}
\begin{tabular}{l|cc|c|cccc}
\multicolumn{1}{c}{} 
& \multicolumn{2}{c}{Fit result} &
  \multicolumn{5}{c}{Parameter  Correlation Coefficients} \\
\hline
& value & error  &  Global  &  $\mc [\GeVcc]$ &   $\kappa$  & $\epsp$ & $\Bmu$ \\
\hline
        $\mc [\GeVcc]$        & 1.258  & 0.160  &  0.74986 &  1.000 &  0.522 & -0.732 &  0.018 \\
        $\kappa$     & 0.326  & 0.048  &  0.56786 &  0.522 &  1.000 & -0.534 & -0.018 \\ 
        $\epsp$      & 0.0646 & 0.0053 &  0.75471 & -0.732 & -0.534 &  1.000 &  0.031 \\
        $\Bmu$      & 0.0959 & 0.0038 &  0.06935 &  0.018 & -0.018 &  0.031 &  1.000 \\
\end{tabular}
\end{center}
\end{table}

A Monte-Carlo reference sample has been used to avoid the painful
procedure in which the likelihood is evaluated by generating a different sample of events 
for each set of parameters.
 The charm quark mass was set to $\mc^{\rm ref}=0.5\ \GeVcc$
to avoid a possible bias due to the vanishing charm production cross-section
below the $\mc^{\rm ref}$ value, while the other parameters were set to $\kappa^{\rm ref}=0.45$ 
and $\epsp^{\rm ref}=0.07$. All Monte-Carlo ``true'' information has been 
retained to be able to weight each event  
when any of the four parameters  is changed during the fitting
procedure.
The reference sample consists of 17439 events that survived the selection criteria out 
of $3.5\times 10^5$  fully reconstructed generated events.

\begin{figure}
\centering \epsfig{figure=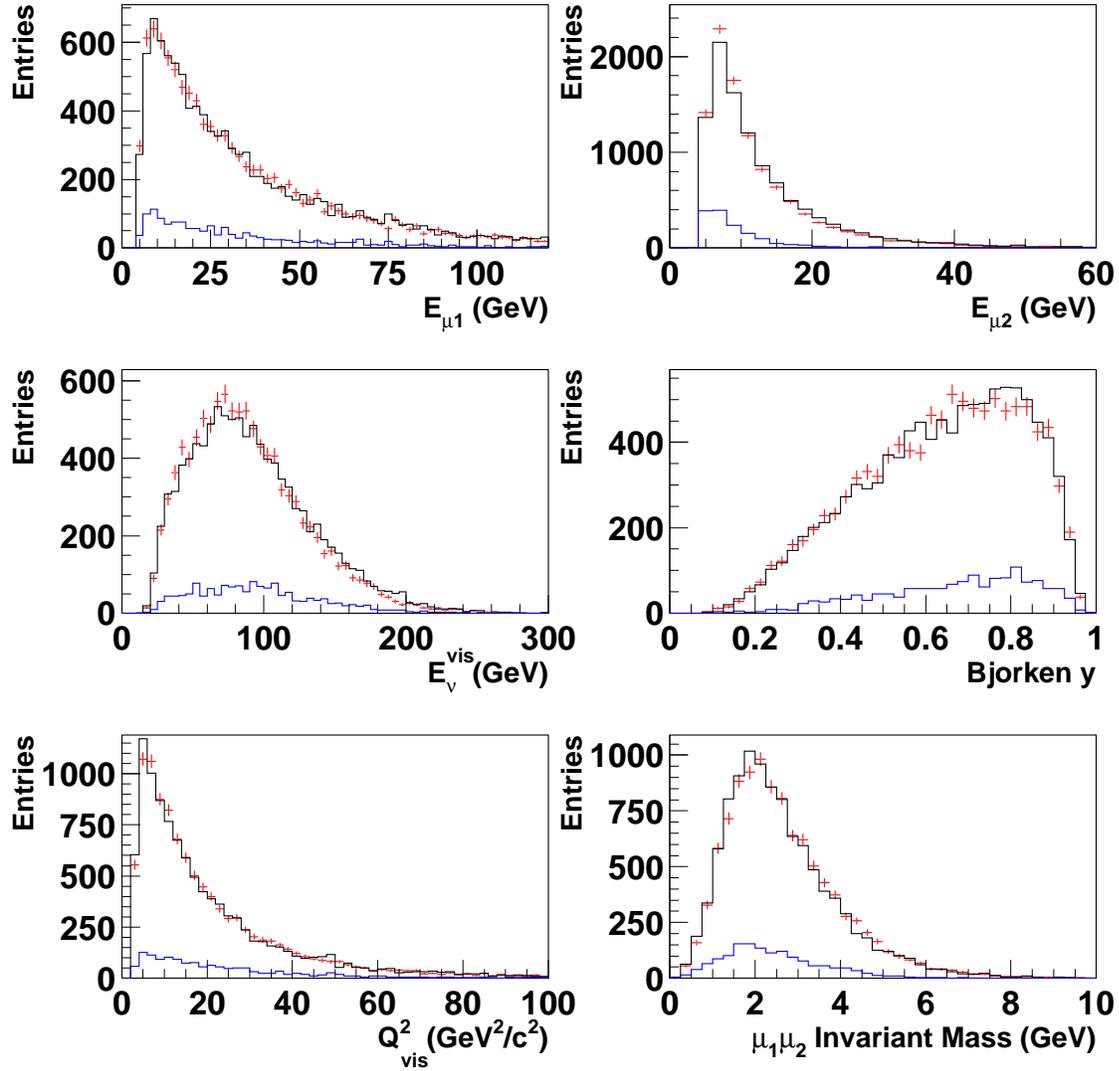,width=1.0\linewidth}
\caption{\label{datamc1} Comparison between data (points) and Monte-Carlo (histogram) evaluated
using the parameters found by the fit procedure. 
The small fraction of events originating from pions and kaons decaying into muons is also shown.}
\end{figure}

\begin{figure}
\centering \epsfig{figure=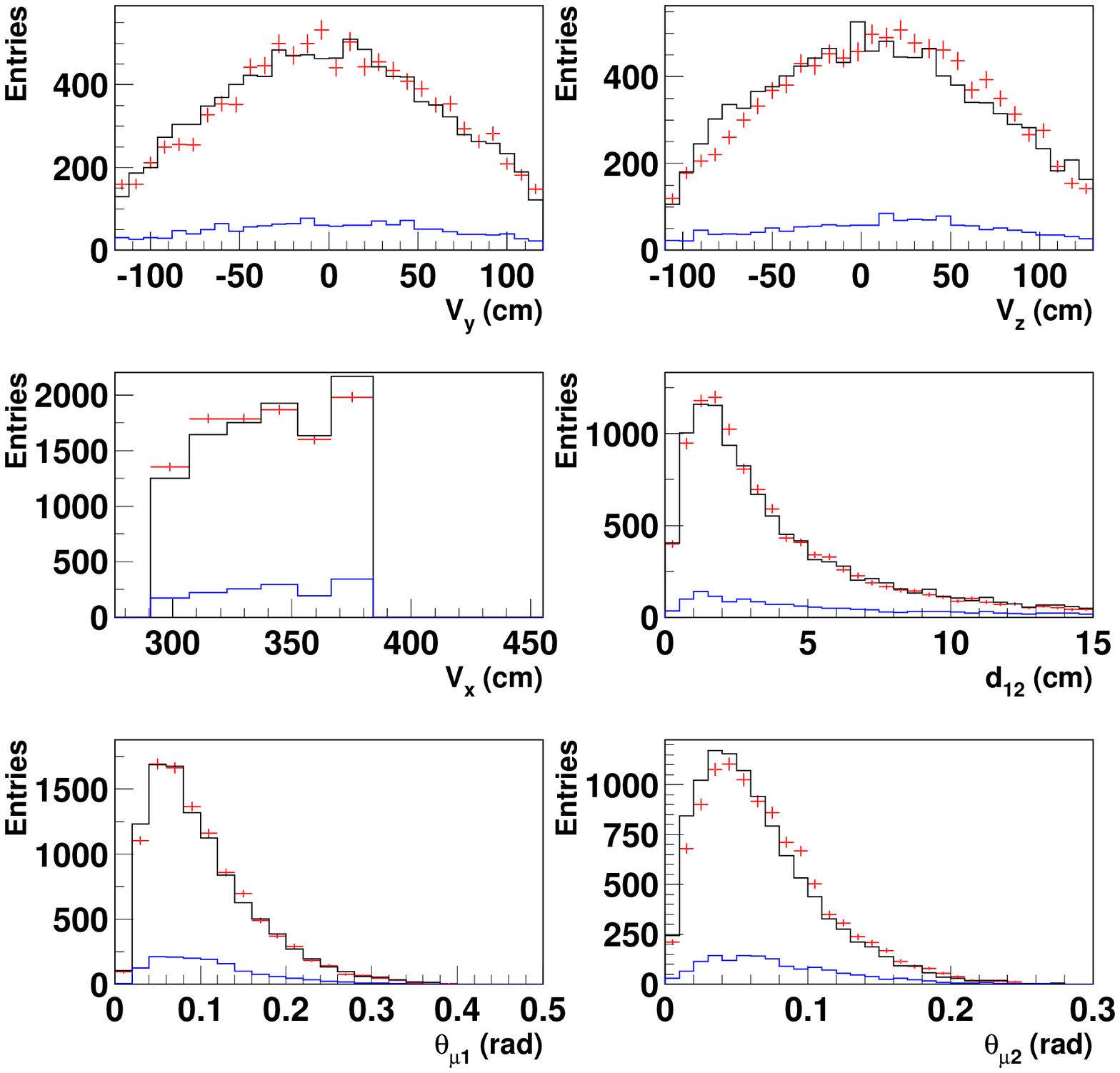,width=1.0\linewidth}
\caption{\label{datamc2} Comparison between data (points) and Monte-Carlo (histogram) evaluated
using the parameters found by the fit procedure.
The small fraction of events originating from pions and kaons decaying into muon is also shown.}
\end{figure}

An unbinned likelihood fit of the four unknown parameters has been performed on the basis
of the distributions of the three experimental observables
$E_\nu^{\rm vis}$, $x_{\rm vis}$ and $z_{\rm vis}$. The probability
density functions of the three observables are constructed by assigning 
to each event in the reference sample a weight equal to the ratio 
of the cross-section evaluated with the current set of parameters to the cross-section
obtained with the reference set. 
The absolute normalization of the Monte-Carlo sample was obtained
by using the observed dimuon production rate with respect to the single muon 
events and fixing the absolute value of the cross-section with the same
method as reported in Ref.~\cite{sfrolf}. 

The fit was performed using MINUIT~\cite{minuit} simultaneously on the
four unknown parameters and 
the result of the fit procedure is reported in Table~\ref{fitres}.
Figures~\ref{datamc1} and \ref{datamc2} show the comparison between the data and 
the Monte-Carlo samples produced using the best fit values 
reported in Table~\ref{fitres}.
The result of the fit is very stable and consistent values are obtained by 
fitting separately each parameter keeping the others fixed. 

\subsection{Systematic uncertainties}

Systematic uncertainties on the result of the fit are mainly due to
the theoretical model, the background
description and the neutrino flux normalization. 
Theoretical uncertainties were taken as the maximum deviation of each fitted value
from its central value when varying the parameters of the model inside their quoted errors
and repeating the fit procedure. The contribution of the nucleon {\em pdf}\, has been evaluated
repeating the fit procedure using the CTEQ3L {\em pdf}\, parametrization. The amount of events
induced by pion and kaon decays, the background subtraction procedure and the total 
uncertainty on the overall event normalization set the limit on the precision of 
$\Bmu$ and of the ratio of the dimuon to the charged-current cross-sections
$\sigma_{-+}/\sigma_\mathrm{CC}$.
The systematic errors due to the fiducial volume and the event selection criteria 
definition have been evaluated by
varying these cuts inside the resolutions of the corresponding variables.
The major contributions to the total systematic errors
on the four fitted parameters are shown in Table~\ref{syst}. 
\begin{table}
\caption{\label{syst} Major sources of systematic error uncertainties to the evaluated values 
of $\mc$, $\kappa$ and $\epsp$ and $\Bmu$.}
\begin{center}
\begin{small}
\begin{tabular}{l cccc}
 &  $\Delta \mc\ [\GeVcc]$  &   $\Delta \kappa$  &  $\Delta \epsp$  &  $\Delta \Bmu$ \\
\hline
$\mc\ [\GeVcc]$                       &  $-$   &  0.02   &   0.004   &  0.001   \\
$\kappa$                            &  0.02  &  $-$    &   0.005   &  0.001   \\
$\epsp$                          &  0.03  &  0.02   &   $-$     &  0.001   \\
Event normalization                 &  $-$   &  $-$    &   $-$     &  0.003   \\
Background scale                    &  0.04  &  0.02   &   0.004   &  0.005   \\
Fit procedure                       &  0.01  &  0.002  &   0.001   &  0.0005  \\
$\Esh$ scale (5\%)                  &  0.03  &  0.013  &   0.001   &  0.004   \\
$\Esh$ offset ($\pm$ 2 GeV)         &  0.05  &  0.011  &   0.001   &  0.001  \\
Radiative corrections               &  0.02  &  0.005  &   0.002   &  0.0005  \\
$V_\mathrm{cd}$, $V_\mathrm{cs}$, nucleon {\em pdf}   &  0.02  &  0.015  &   0.001   &  0.002   \\
Fiducial volume                     &  0.01  &  0.02   &   0.001   &  0.0012  \\
Selection cuts                      &  0.02  &  0.003  &   0.002   &  0.0011  \\
\hline
Total                         &     0.088    &  0.046  &   0.0084  &   0.0078 \\
\end{tabular}
\end{small}
\end{center}
\end{table}

Including the systematic uncertainties,
the result of this leading order analysis can be summarized as
follows:
\begin{eqnarray}
 \mc &=&    (1.26  \pm \ 0.16  {\rm (stat)} \pm \ 0.09  {\rm (syst))\ \GeVcc} \nonumber \\
 \kappa &=&    0.33  \pm \ 0.05  {\rm (stat)} \pm \ 0.05  {\rm (syst)} \\
 \epsp &=&    0.065 \pm \ 0.005 {\rm (stat)} \pm \ 0.009 {\rm (syst)} \nonumber \\
 \Bmu &=&    0.096 \pm \ 0.004 {\rm (stat)} \pm \ 0.008 {\rm (syst)} \nonumber
\end{eqnarray}

\section{Discussion and conclusions}

The results of this analysis are in agreement with those reported by 
previous experiments~\cite{cdhs,ccfr,charmii,nomad,nutev}, as shown in Table~\ref{comp}. 
\begin{table}[b]
\caption{\label{comp} Compilation of latest results on neutrino induced opposite-sign dimuon events.}
\begin{center}
\begin{small}
\begin{tabular}{lrrcccc}
Experiment & $N_{2\mu}(\nu)$ & $N_{2\mu}(\overline{\nu})$ & $\mc\ [\GeVcc]$ & $\kappa$ & $\Bmu$ & $\epsp$ \\
\hline
This analysis   &  8910 &  430 & $1.26 \pm 0.18$  & $0.33 \pm 0.07$   & $0.096 \pm 0.008$ & $0.065 \pm 0.010$  \nonumber    \\
CDHS~\cite{cdhs}     &  9922 & 2123 &          $-$     & $0.47 \pm 0.09$   & $0.084 \pm 0.014$ & $\left[0.02,0.14\right]$  \nonumber  \\ 
CCFR~\cite{ccfr}     &  4503 &  632 & $1.3 \pm 0.2$    & $0.44 \pm 0.09$   & $0.109 \pm 0.010$ & $-$                 \nonumber    \\
CHARM II~\cite{charmii} &  3100 &  700 & $1.8 \pm 0.4$    & $0.39 \pm 0.09$   & $0.091 \pm 0.010$ & $0.072 \pm 0.017$  \nonumber    \\
NOMAD~\cite{nomad}    &  2714 &  115 & $1.3 \pm 0.4$    & $0.48 \pm 0.17$   & $0.095 \pm 0.015$ & $0.08  \pm 0.05$       \\
NuTeV~\cite{nutev}    &  2280 &  655 & $1.3 \pm 0.2$    & $0.38 \pm 0.08$   & $0.101 \pm 0.012$ & $-$                \nonumber    
\end{tabular}
\end{small}
\end{center}
\end{table}

The value of $\Bmu|V_\mathrm{cd}|^2$ obtained using the results of this analysis can be 
combined directly with other leading order results~\cite{cdhs,ccfr,charmii} following the 
prescription given in Ref.~\cite{pmiglioz}. 
An average value of
\begin{equation} 
(\Bmu|V_\mathrm{cd}|^2)_{\rm LO} = (0.474 \pm 0.027)\times 10^{-2}
\end{equation}
is obtained.

Results of this analysis may also be compared with earlier
analyses of events originating in the nuclear emulsion target of the
CHORUS experiment.
Good agreement is found with the result $\Bmu = 0.085\pm 0.010$ 
obtained after applying the selection $E_{\nu} > 30 \ \GeVc$~\cite{sergei}.
The use of that selection for the comparison is justified by the small
contribution of events below $30~\GeV$ in the present
analysis, as visible in Fig.~\ref{datamc1}.
The value of the Peterson fragmentation parameter 
$\epsp = 0.059\pm 0.013$ reported in Ref.~\cite{plb604} for the same
definition 
of $z$ as used here (called $\epsp^Q$ in Ref.~\cite{plb604}) is in good 
agreement with the
result found in the present analysis.  
In Ref.~\cite{plb613}, the value $\mc = (1.42 \pm 0.08)\ \GeVcc$ was reported,
assuming $\kappa = 0.38$ and
$\alpha = 1$. 
The dependence of $\mc$ on $\kappa$ and $\alpha$ is given in Ref.~\cite{plb604}.
For  $\alpha = 2$, as used here, and fixing $\kappa$ to its fitted value 
$0.33$, the value of $\mc$ becomes $1.30 \pm 0.08\ \GeVcc$ in excellent agreement with
the value found in the present analysis.

\begin{figure}
\centering \epsfig{figure=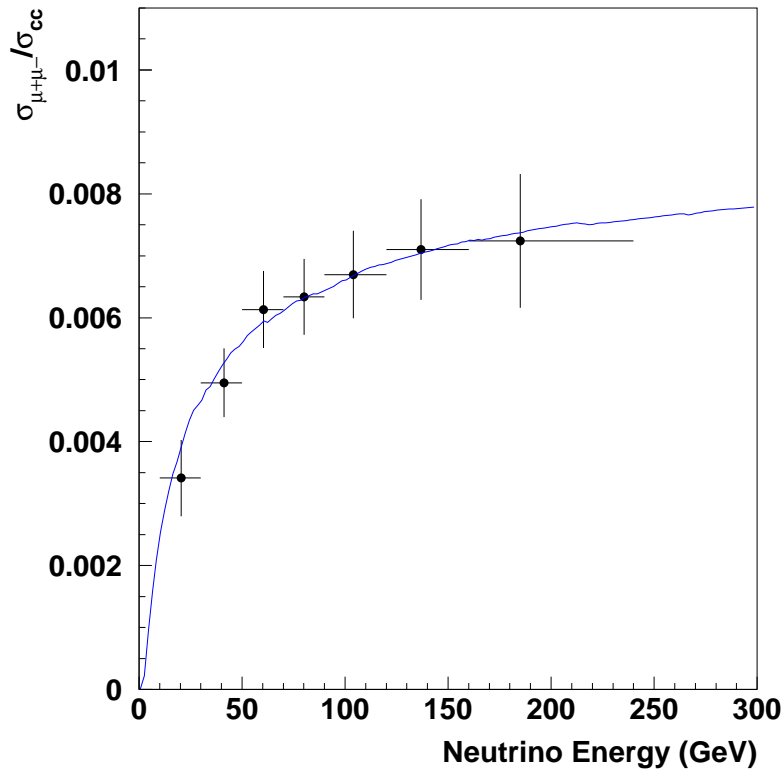,width=0.7\linewidth}
\caption{\label{eplot} Opposite-sign dimuon production relative cross-section.
The solid line represents the Monte-Carlo prediction obtained using the values of Table~\ref{fitres}.}
\end{figure}

As an additional result, the large sample of collected dimuon events
allowed us 
to make the evaluation of the rate of charm induced 
opposite-sign dimuon events relative to the charged-current events 
as a function of the incoming neutrino energy. 
An unfolding procedure has been used to take into account
all detector effects. The correlation between different energy bins has been evaluated using 
a Monte-Carlo simulation. The correction due to the missing energy of the outgoing neutrino 
in the charm decay has been calculated for each energy bin using the Monte-Carlo event sample. 
The result is shown in Fig.~\ref{eplot} where statistical and systematic errors have been 
added in quadrature. It agrees well with the world average dimuon rate shown in Fig.$26$ of
Ref.~\cite{pmiglioz}.

In summary, the analysis of the second largest sample of neutrino induced dimuon events
has been performed using a leading order QCD formalism. The slow rescaling mechanism gives 
a good description of the CHORUS data. Results on the charm quark mass $\mc$, the content of 
strange quark in the nucleon $\kappa$, the Peterson fragmentation parameter $\epsp$ and the 
branching fraction of charmed hadrons into muon, $\Bmu$, agree with the data obtained in similar 
analyses and improve the overall knowledge on the neutrino induced charm production and charm 
decay mechanisms.

\section{Acknowledgments}

We gratefully acknowledge the help and support of the neutrino beam staff and of the numerous technical
collaborators who contributed to the detector construction and operation. The experiment was made possible
by grants from the Institut Interuniversitaire des Sciences Nucl\'eaires and the Interuniversitair 
Instituut voor Kernwetenschappen (Belgium), the Israel Science Fundation (grant 328/94) and the Technion
Vice President Fund for the Promotion of Research (Israel), CERN (Geneve, Switzerland), the German 
Bundesministerium f\"ur Bildung und Forschung (Germany), the Institute of Theoretical and Experimental
Physics (Moscow, Russia), the Istituto Nazionale di Fisica Nucleare (Italy), the Promotion and Mutual
Aid Corporation for Private Schools of Japan and Japan Society for the Promotion of Science (Japan), the 
Korea Research Foundation Grant (KRF-2003-005-C00014) (Republic of Korea), the Foundation for Fundamental 
Research Organization NWO (The Netherlands), and the Scientific and Technical Research Council of Turkey
(Turkey). We gratefully acknowledge their support.


\begin{thebibliography}{0}

\bibitem{cdhs} H.~Abramowicz {\it et al.}, Z. Phys. {\bf C15} (1982) 19.

\bibitem{ccfr} S.A.~Rabinowitz {\it et al.}, Phys. Rev. Lett. {\bf 70} (1993) 134; \\
               A.O.~Bazarko {\it et al.}, Z. Phys. {\bf C65} (1995) 189.

\bibitem{charmii} P.~Vilain {\it et al.}, Eur. Phys. J. {\bf C11} (1999) 19-34.

\bibitem{nomad} P.~Astier {\it et al.}, Phys. Lett. {\bf B486} (2000) 35-48.

\bibitem{nutev} M.~Goncharov {\it et al.}, Phys. Rev. {\bf D64} (2001) 112006.

\bibitem{bebc} G.~Gerbier {\it et al.}, Z. Phys. {\bf C29} (1985) 15.

\bibitem{e531} N.~Ushida {\it et al.}, Phys. Lett. {\bf B121} (1983) 292.

\bibitem{plb604} G.~\"Onengut {\it et al.}, Phys. Lett. {\bf B604} (2004) 145-156.

\bibitem{plb613} G.~\"Onengut {\it et al.}, Phys. Lett. {\bf B613} (2005) 105-117.

\bibitem{sergei} A.~Kayis-Topaksu {\it et al.}, Phys. Lett. {\bf B626} (2005) 24-34.

\bibitem{chorus} E.~Eskut {\it et al.}, Phys. Lett. {\bf B497} (2001) 8-22.

\bibitem{chorusdet} E.~Eskut {\it et al.}, Nucl. Inst. and Methods {\bf A401} (1997) 7.

\bibitem{calor} E.~Di Capua {\it et al.}, Nucl. Inst. and Methods {\bf A378} (1996) 221.

\bibitem{trigger} M.G.~van Beuzekom {\it et al.}, Nucl. Inst. and Methods {\bf A427} (1999) 587.

\bibitem{trimu} A.~Kayis-Topaksu {\it et al.}, Phys. Lett. {\bf B596} (2004) 44-53.

\bibitem{slowr} H.~Georgi and H.D.~Politzer {\it et al.}, Phys. Rev. {\bf D14} (1976) 1829.

\bibitem{aivazis} M.A.G.~Aivazis {\it et al.}, Phys. Rev. {\bf D50} (1994) 3085;
                  M.A.G.~Aivazis {\it et al.}, Phys. Rev. {\bf D50} (1994) 3102.

\bibitem{barnett} R.M.~Barnett, Phys. Rev. Lett. {\bf 36}  (1976) 1163.

\bibitem{bardin} D.Yu.~Bardin and V.A.~Dokuchaeva, Preprint JINR E2-86-260 (1986).

\bibitem{grv94lo} M.~Gl\"uck, E.~Reya and A.~Vogt, Z. Phys. {\bf C67} (1995) 433-448.

\bibitem{peterson} C.~Peterson {\it et al.}, Phys. Rev. {\bf D27} (1983) 105.

\bibitem{nlo} M.~Gl\"uck {\it et al.}, Phys. Lett. {\bf B398} (1997) 391.

\bibitem{lebc} M.~Aguilar-Benitez {\it et al.}, Phys. Lett. {\bf B123} (1983) 98.

\bibitem{jetset} T.~Sjostrand, Comput. Phys. Commun. {\bf 82} (1994) 74.

\bibitem{geant} GEANT 3.21, CERN program library long writeup W5013.

\bibitem{pdflib} H.~Plothow-Besch, PDFLIB, CERN program library long writeup W5051.

\bibitem{cteq3l} H.L.~Lai {\it et al.}, Phys. Rev. {\bf D51} (1995) 4763.

\bibitem{pdg} S.~Eidelman {\it et al.}, Phys. Lett. {\bf B592} (2004) 1.

\bibitem{sfrolf} G.~\"Oneng\"ut {\it et al.}, Phys. Lett. {\bf B632} (2006) 65-75.

\bibitem{minuit} F.~James, MINUIT:Function Minimization and Error Analysis, CERN program library entry D506.

\bibitem{pmiglioz} P.~Migliozzi {\it et al.}, Phys. Rep. {\bf 399} (2004) 227-320.

\end{thebibliography}
\end{document}